\documentclass[Physsubmission, Phys]{SciPost}
\hypersetup{
    colorlinks,
    linkcolor={red!50!black},
    citecolor={blue!50!black},
    urlcolor={blue!80!black}
}

\usepackage[bitstream-charter]{mathdesign}
\DeclareSymbolFont{usualmathcal}{OMS}{cmsy}{m}{n}
\DeclareSymbolFontAlphabet{\mathcal}{usualmathcal}

\begin{document}

% TODO: write your article's title here.
% The article title is centered, Large boldface, and should fit in two lines
\begin{center}{\Large \textbf{
Hidden dependencies in model independent tests of DAMA\\
}}\end{center}

% TODO: write the author list here. Use initials + surname format.
% Separate subsequent authors by a comma, omit comma at the end of the list.
% Mark the corresponding author with a superscript *.
\begin{center}
Madeleine J. Zurowski\textsuperscript{1,2$\star$}
\end{center}

% TODO: write all affiliations here.
% Format: institute, city, country
\begin{center}
{\bf 1} School of Physics, The University of Melbourne, Melbourne, VIC 3010, Australia
\\
{\bf 2} ARC Centre of Excellence for Dark Matter Particle Physics, Australia
\\
% TODO: provide email address of corresponding author
* madeleine.zurowski@unimelb.edu.au
\end{center}

\begin{center}
\today
\end{center}

% For convenience during refereeing (optional),
% you can turn on line numbers by uncommenting the next line:
%\linenumbers
% You should run LaTeX twice in order for the line numbers to appear.

\definecolor{palegray}{gray}{0.95}
\begin{center}
\colorbox{palegray}{
  \begin{tabular}{rr}
  \begin{minipage}{0.1\textwidth}
    \includegraphics[width=30mm]{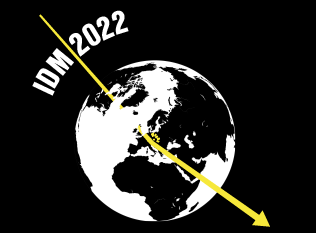}
  \end{minipage}
  &
  \begin{minipage}{0.85\textwidth}
    \begin{center}
    {\it 14th International Conference on Identification of Dark Matter}\\
    {\it Vienna, Austria, 18-22 July 2022} \\
    \doi{10.21468/SciPostPhysProc.?}\\
    \end{center}
  \end{minipage}
\end{tabular}
}
\end{center}

\section*{Abstract}
{\bf
% TODO: write your abstract here.
For nearly two decades the DAMA Collaboration has been observing a modulating signal compatible with that expected from a dark matter presence in our galaxy. However, interpretations of this with the standard assumptions for dark matter particles are strongly ruled out by a large number of other experiments. This tension can be relaxed somewhat by making more tailored choices for the dark matter model and properties of interest, but expanding the models of interest in such a way makes it impossible to test the DAMA modulation conclusively. In order to understand the exact nature of this signal, we need to use a detector based on the same target (NaI(Tl)), which would be sensitive to exactly the same particle interaction models as DAMA.
There are a number of such experiments in the data taking or commissioning stages designed to do just this, two of which (ANAIS and COSINE) recently released their results after 3 years of data taking. Interestingly, the modulation observed by the two experiments deviate from each other by 2$\sigma$, while being within 3$\sigma$ of the DAMA result. This paper addresses potential differences between NaI(Tl) based detectors that could lead to the differing results to date, with a particular focus on the quenching factor
}

% TODO: include a table of contents (optional)
% Guideline: if your paper is longer that 6 pages, include a TOC
% To remove the TOC, simply cut the following block
% \vspace{10pt}
% \noindent\rule{\textwidth}{1pt}
% \tableofcontents\thispagestyle{fancy}
% \noindent\rule{\textwidth}{1pt}
% \vspace{10pt}

\section{Introduction}
\label{sec:intro}
Dark matter (DM) has long been postulated as a solution to a number of astrophysical observations on a large range of scales \cite{bertone}. While a variety of different detection methods have been pursued across direct, indirect, and collider searches, to date there is no clear signal that the community can reach a consensus on having a DM origin. Despite this, there have been a number of observations, or anomalies, that at face value seem compatible with DM. One that has received significant attention over the past decade are the results from the DAMA collaboration, which report a signal that modulates over the year with a significance of 12.9$\sigma$ \cite{BERNABEI2020}. For DM, such a signal is due to the relative motion of the Earth through the distribution of DM within the galaxy. This is expected to produce a modulation with a period of 1 year and a peak in June, which matches the DAMA observation. However, this result is incompatible with null results from almost every other DM direct detection experiment under typically assumed interaction models \cite{Patrignani:2016xqp}. Although this tension suggests that this modulation is from something other than DM, a truly conclusive test requires an experiment that uses the same target as DAMA, NaI(Tl), to be sensitive to exactly the same interactions. A number of such experiments are planned, with two already taking data, and initial results have yet to completely rule out the DAMA modulation.\\
In the past year, both ANAIS and COSINE have released three years of data analysis for their modulation searches \cite{Adhikari2021,ANAIS2021}. Although error bars are at present still large, there appears to be a discrepancy between the three experiments, despite the fact that they use the same target. It is possible that differences between these detectors, despite the fact that they use the same target, can introduce a `hidden' model dependence - that is that changes may appear more extreme for different models or mass of DM. One possible difference between the setups that could explain this is the quenching factor.

% \begin{figure}[!h]
%     \centering
%     \includegraphics[width=\textwidth]{indep-res.png}
%     \caption{Annual modulation reported by DAMA, ANAIS, and COSINE, along with the naive weighted global average.}
%     \label{fig:indep-res}
% \end{figure}

\section{Quenching factor}
\begin{figure}[!h]
    \centering
    \includegraphics[width=0.75\textwidth]{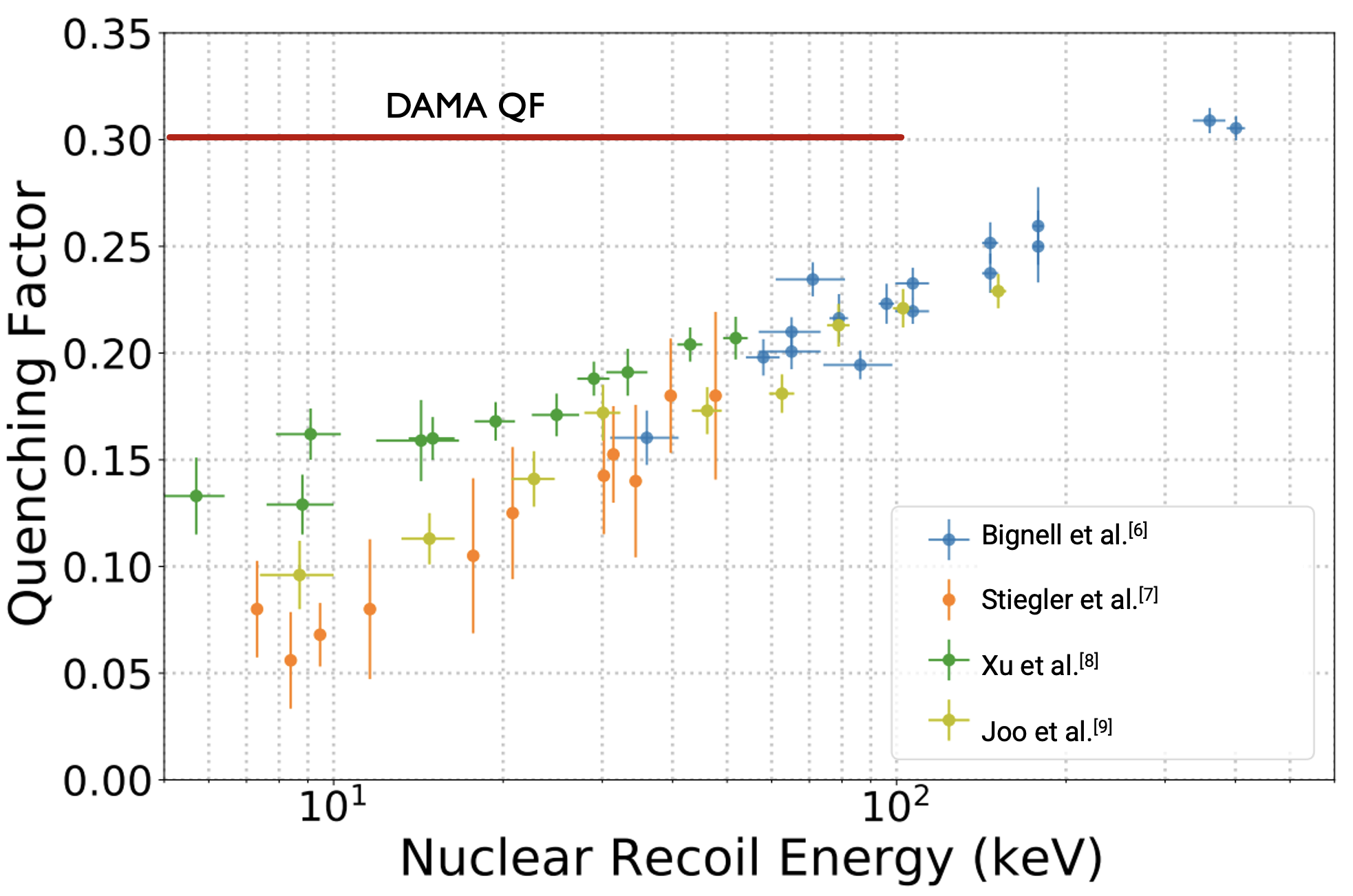}
    \caption{Na QF measured by different groups, adapted from Ref. \cite{bignell2021}, and including results from Refs. \cite{stiegler2017,Xu_2015,JOO201950}.}
    \label{fig:qf-vals}
\end{figure}

The purpose of a quenching factor (QF) in NaI(Tl) experiments is to convert the nuclear recoil energy $E_{NR}$ (the signal of interest) into electron equivalent energy $E_{\rm ee}$ (the units of the detector)
\begin{equation}
    E_{\rm ee} = Q(E_{NR})E_{NR}.
\end{equation}
This is required because not all of the energy imparted into the recoiling nucleus is transformed into a detectable signal. It is possible that this effect depends strongly on the optical properties of the crystal, and if so then different growth methods can impact results. A variable QF is interesting to consider as there have been a number of different measurements across various groups, shown in Fig. \ref{fig:qf-vals}, and because it would mean that both the amplitude and energy of any signal would be expected to change. As NaI(Tl) is a composite target, it will effect different DM masses in different ways. For example, DM will interact preferentially with targets of a similar mass, so if it is low mass and interacts with Na, a varying the I QF will have very little impact, and vice versa. Because of this, correcting for different QFs cannot take the form of a simple scale factor.\\

\begin{figure}[!h]
    \centering
    \includegraphics[width=0.9\textwidth]{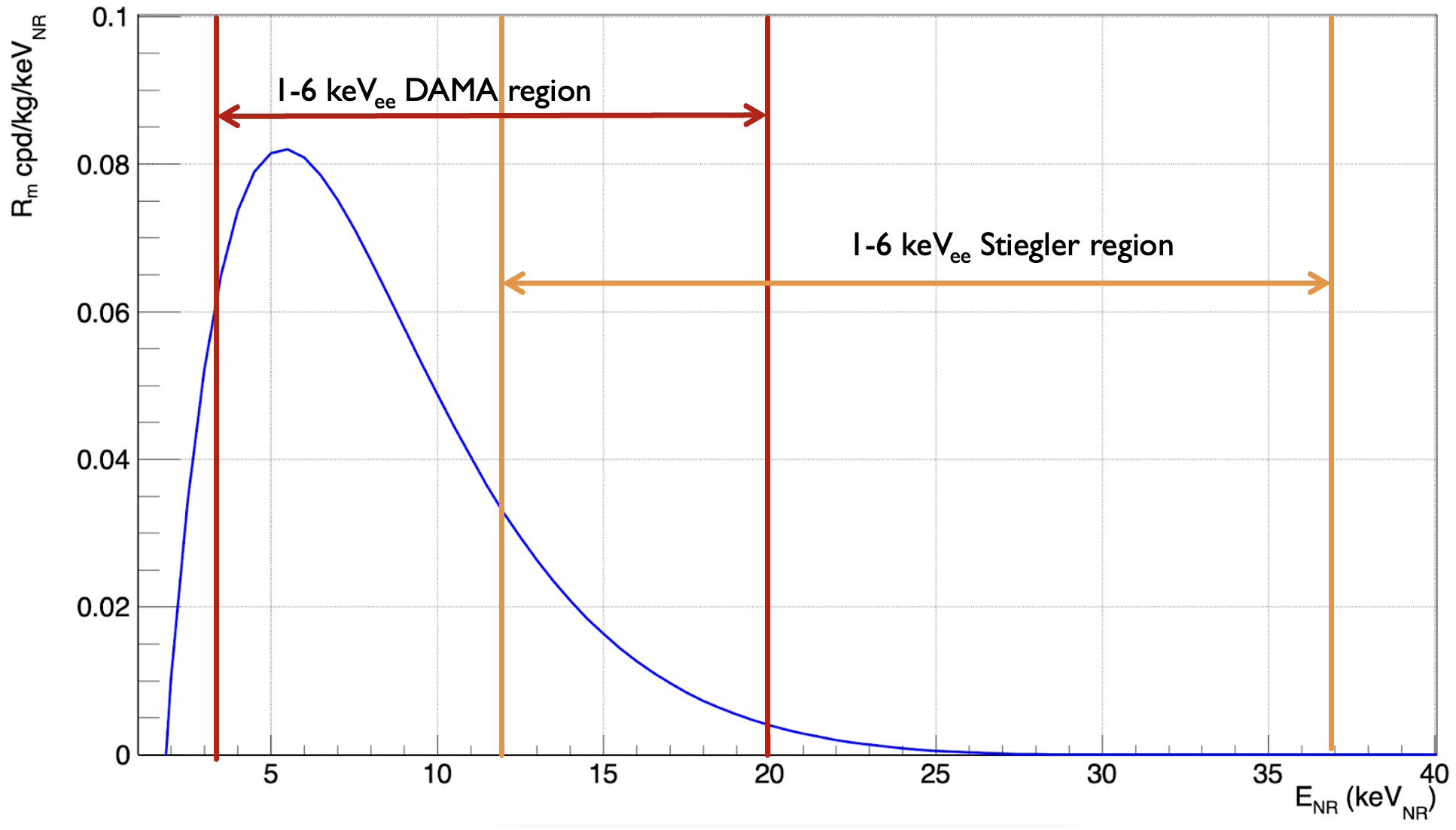}
    \caption{Recoil energy spectrum of an illustrative DM model. Superimposed in red and orange are the regions that correspond to 1-6 keV$_{\rm ee}$ after the application of either the DAMA (red) or Stiegler (orange) QFs.}
    \label{fig:qf-regions}
\end{figure}

Crystals that follow different QF models will have a region of interest (defined for NaI(Tl) detectors at 1-6 keV$_{\rm ee}$) that accesses different parts of the recoil energy spectrum. An example of this for a typical DM recoil energy spectrum is given in Fig. \ref{fig:qf-regions}, where a DAMA-like QF model will have a clear peak occurring in the region of interest, while a Stiegler-like QF model will only observe the exponential tail. This effect will impact any nuclear recoil signal in a NaI(Tl) detector, not just DM, where the extremity of the impact is dictated by the overall shape and features of the nuclear recoil energy spectrum.

\begin{figure}[!h]
    \centering
    \includegraphics[width=\textwidth]{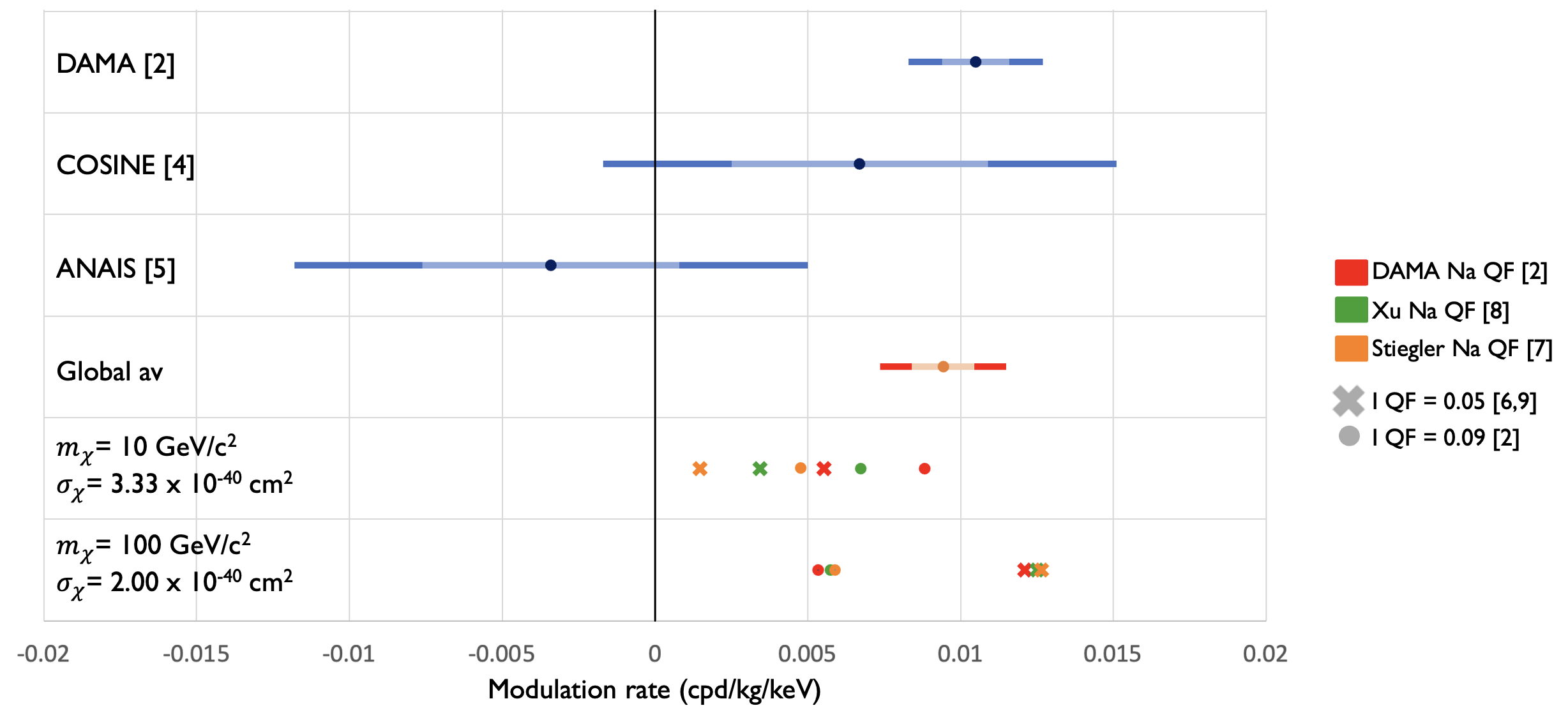}
    \caption{The modulation rate observed in the 1-6 keV region for the various operational experiment compared with the expected modulation for a 10 or 100 GeV/c$^2$ WIMP under the assumption of various QFs.}
    \label{fig:qf-impact}
\end{figure}

The consideration that different NaI(Tl) crystals have different QFs can help to reduce the apparent tension between the results reported by various NaI(Tl) detectors. As an example, the expected modulation that would be observed for a DM with a mass of 10 GeV/c$^{2}$ assuming a spin independent interaction is compared with the reported results under various combinations of Na and I QFs in Fig. \ref{fig:qf-impact}. Each of the experimental results shown here are within 1$\sigma$ of at least one viable QF combination: if ANAIS crystals have a Na QF given by the Stiegler model, and an I QF of 0.05, while COSINE crystals  have a Xu Na QF and an I QF of 0.09, and DAMA crystals are well modelled by the DAMA QF and I QF of 0.09, then all three experiments are observing a modulation consistent with expectations from this particular DM case.

\section{Impact on model independence}
For a single target (e.g., Ge or Xe), if the QF (or equivalent transformation) is well known and modelled, it can easily be accounted for in comparisons by rescaling the observed rate, and reporting results in nuclear recoil energy rather than electron equivalent. For composite targets like NaI(Tl), however, there is no way of knowing whether an observed event came from the recoil of an I or Na nucleus. The interaction of interest must first be modelled to understand the expected ratio of Na vs I recoil events at various energies, and the appropriate QF applied accordingly. For DM, this requires the assumption of some DM mass and interaction model, and depending on this choice, the effects on the expected rate variance between crystals with different QFs will change. For example, Fig. \ref{fig:qf-impact} compares the impact of different QF assumptions for two different DM masses, assuming a standard spin independent interaction. While the expected rate of the 10 GeV/c$^{2}$ DM is very sensitive to changes in Na QF, the 100 GeV/c$^{2}$ sees almost no change. Because this effect is so dependent on mass, if QF is a crystal-dependent factor, model independent tests become impossible.

\section{Conclusion}
A number of NaI(Tl) detectors designed to provide a model independent test of DAMA are at present observing different modulation rates. Crystal dependent QFs offer an explanation for this, but in doing so introduce a degree of model dependent. If this is the cause for the observed results, truly model independent tests of DAMA become almost impossible. Further studies are required to carefully understand the QFs for the currently operating and planned NaI(Tl) experiments to fully understand this impact. 

\section*{Acknowledgements}
MJZ is funded by an Australian Government Research Training Program Scholarship the Australian Government, and through the Australian Research Council grant CE200100008. The presentation of this work was partially funded by the Laby Foundation.

% TODO:
% Provide your bibliography here. You have two options:

% FIRST OPTION - write your entries here directly, following the example below, including Author(s), Title, Journal Ref. with year in parentheses at the end, followed by the DOI number.
%\begin{thebibliography}{99}
%\bibitem{1931_Bethe_ZP_71} H. A. Bethe, {\it Zur Theorie der Metalle. i. Eigenwerte und Eigenfunktionen der linearen Atomkette}, Zeit. f{\"u}r Phys. {\bf 71}, 205 (1931), \doi{10.1007\%2FBF01341708}.
%\bibitem{arXiv:1108.2700} P. Ginsparg, {\it It was twenty years ago today... }, \url{http://arxiv.org/abs/1108.2700}.
%\end{thebibliography}

% SECOND OPTION:
% Use your bibtex library
% bibliographystyle{SciPost_bibstyle} % Include this style file here only if you are not using our template
\bibliography{bib.bib}

\begin{thebibliography}{1}
\providecommand{\url}[1]{\texttt{#1}}
\providecommand{\urlprefix}{URL }
\expandafter\ifx\csname urlstyle\endcsname\relax
  \providecommand{\doi}[1]{doi:\discretionary{}{}{}#1}\else
  \providecommand{\doi}{doi:\discretionary{}{}{}\begingroup
  \urlstyle{rm}\Url}\fi
\providecommand{\eprint}[2][]{\url{#2}}

\bibitem{bertone}
G.~Bertone and D.~Hooper,
\newblock \emph{{History of dark matter}},
\newblock Rev. Mod. Phys \textbf{90}(045002) (2018).

\bibitem{BERNABEI2020}
R.~Bernabei \emph{et~al.},
\newblock \emph{{The DAMA project: Achievements, implications and
  perspectives}},
\newblock PPNP \textbf{114}, 103810 (2020),
\newblock \doi{https://doi.org/10.1016/j.ppnp.2020.103810}.

\bibitem{Patrignani:2016xqp}
C.~Patrignani \emph{et~al.},
\newblock \emph{{Review of Particle Physics}},
\newblock Chin. Phys. \textbf{C40}(10), 100001 (2016),
\newblock \doi{10.1088/1674-1137/40/10/100001}.

\bibitem{Adhikari2021}
G.~Adhikari \emph{et~al.},
\newblock \emph{{Three-year annual modulation search with COSINE-100}} (2021),
  \eprint{2111.08863}.

\bibitem{ANAIS2021}
J.~Amar\'e \emph{et~al.},
\newblock \emph{{Annual modulation results from three-year exposure of
  ANAIS-112}},
\newblock Phys. Rev. D \textbf{103}, 102005 (2021),
\newblock \doi{10.1103/PhysRevD.103.102005}.

\bibitem{bignell2021}
L.~J. Bignell, I.~Mahmood \emph{et~al.},
\newblock \emph{Quenching factor measurements of sodium nuclear recoils in
  {NaI:Tl} determined by spectrum fitting},
\newblock preprint (2021), \eprint{2102.02833}.

\bibitem{stiegler2017}
T.~Stiegler \emph{et~al.},
\newblock \emph{A study of the {NaI(Tl)} detector response to low energy
  nuclear recoils and a measurement of the quenching factor in {NaI(Tl)}},
\newblock preprint (2017), \eprint{1706.07494}.

\bibitem{Xu_2015}
J.~Xu \emph{et~al.},
\newblock \emph{Scintillation efficiency measurement of na recoils in {NaI(Tl)}
  below the dama/libra energy threshold},
\newblock Physical Review C \textbf{92}(1) (2015),
\newblock \doi{10.1103/physrevc.92.015807}.

\bibitem{JOO201950}
H.~Joo, H.~Park, J.~Kim, J.~Lee, S.~Kim, Y.~Kim, H.~Lee and S.~Kim,
\newblock \emph{Quenching factor measurement for {NaI(Tl)} scintillation
  crystal},
\newblock Astroparticle Physics \textbf{108}, 50 (2019),
\newblock \doi{https://doi.org/10.1016/j.astropartphys.2019.01.001}.

\end{thebibliography}

\nolinenumbers

\end{document}